# Scintillations of optical vortex in randomly inhomogeneous medium

Valerii P. Aksenov,* and Valeriy V. Kolosov

*V.E. Zuev Institute of Atmospheric Optics, Russian Academy of Sciences, Siberian Branch*
*1 Academician Zuev square, Tomsk, 634021 Russia*
*\*Corresponding author: avp@iao.ru*

The comparative numerical and analytical analysis of scintillation indices of the vortex Laguerre-Gaussian beam and the и non-vortex doughnut hole and Gaussian beams propagating in the randomly inhomogeneous atmosphere has been performed. It has been found that the dependence of the scintillation index at the axis of the optical vortex on the turbulence intensity at the path has the form of a unit step. It has been shown that the behavior of scintillations in the cross sections of vortex and non-vortex beams differs widely.

Laser beams having the orbital angular momentum (OAM) [1-3] attract recently great attention owing to their particular properties, which have found numerous applications [4-6]. Due to the presence of the transverse circulation component of the Pointing vector [1], such beams are referred to as optical vortices. In particular, the possibility of using optical vortices for information coding and transmission is studied intensely [7-8]. For optical communication systems, it is necessary to study the influence of a medium on the optical vortex propagation. It is known that the medium, the beam propagates through, distorts the beam. Beam wandering and fluctuations of the beam intensity are the main factors restricting the data throughput of optical systems for data transmission along horizontal and slant paths, in particular, Earth-space paths. Intensity fluctuations of Gaussian laser beams in the turbulent atmosphere are studied quite thoroughly now.

To describe fluctuation characteristics of these beams, numerical and analytical methods have been developed [9]. A common feature of an optical vortex in the free space is that the beam structure includes the helical phase distribution and zero intensity at the beam axis [10]. For analytical description of the propagation of laser beams under conditions of weak atmospheric turbulence, the Rytov method is used most often [9]. It was shown in [11-12] that the direct application of the Rytov method for description of vortex beams gives rise to serious problems, because the intensity at the axis of the beam propagating in the undisturbed medium becomes zero. The numerical simulation of laser beam propagation is usually based on the Monte Carlo technique with the use of phase screens [13]. However, the numerically calculated values of intensity fluctuations of vortex beams do not allow us to judge unambiguously the influence of the energy circulation in the beam on the intensity fluctuations in the beam cross section.

Thus, it follows from the results of [14] that intensity fluctuations in the Laguerre-Gaussian beam differ only slightly from intensity fluctuations in the Gaussian beam. The similar conclusion, except for some details, can also be drawn from [15]. From [16 - 17], it follows that under conditions of weak turbulence the intensity fluctuations of laser beam close to the Laguerre-Gaussian one appear to be much stronger than those of the Gaussian beam. In [18], the intensity fluctuations of the vortex Bessel beam behave qualitatively in the same manner as the fluctuations of the Bessel beam having no vortex properties, but their scintillation index appears to be higher. The results [14-18] were obtained with computational grids having different dimensions and at the different number of realizations used for the calculation of the scintillation index. To draw an unambiguous conclusion about the dependence of the scintillation level on the beam type, we perform the numerical simulation of the propagation of different-type beams: Gaussian beam, Laguerre–Gaussian beam $LG_0^1$, and doughnut hole beam (*DH*) in the turbulent atmosphere[19]. Then the numerical results are compared with the asymptotic estimation of scintillation at the beam axis.

We use the following representation of the complex amplitude of the field in the initial plane ($z = 0$)

$$u(r,\varphi,0) = \frac{4}{a}\sqrt{\frac{\Phi}{c}}\left(\frac{\sqrt{2}r}{a}\right)^p \exp\left\{-\frac{r^2}{a^2}\right\}\exp\{il\varphi\}, \qquad (1)$$

where $\{r,\varphi,z\}$ are cylindrical coordinates, $\Phi$ is the total energy flux, $a$ is the initial radius of the Gaussian source, $c$ is the speed of light. If we take $p=1$ and $l=1$, then Eq. (1) describes the circular mode of the Laguerre-Gaussian beam $LG_0^1$. If $p=1$ and $l=0$, then Eq. (1) corresponds to the doughnut hole beam. If both $p=0$ and $l=0$, then the beam in the initial plane takes the form of the Gaussian beam.

To develop the numerical model, we, as in [20-21], apply the method of splitting by physical factors with separation of the diffraction and refraction components of the parabolic wave equation describing the propagation of optical radiation. Diffraction is calculated with the use of the Fast Fourier Transform (FFT) algorithm at the two-dimensional 512×512 grid. All the transformations associated with refraction correspond to the radiation propagation through a pseudo-random phase screen, whose statistics satisfies the conditions of atmospheric turbulence.

In the calculations, the spectrum $\Phi_n(\boldsymbol{\kappa})$ is taken in the form [9]:

$$\Phi_n(\boldsymbol{\kappa}_\perp,0) = 0.033 C_n^2 \frac{\exp(-\kappa_\perp^2/\kappa_a^2)}{\left(\kappa_\perp^2 + \kappa_0^2\right)^{11/6}}\left[1 + 1.802\frac{\kappa_\perp}{\kappa_a} - 0.254\left(\frac{\kappa_\perp}{\kappa_a}\right)^{7/6}\right], \qquad (2)$$

where $C_n^2$ is the structure characteristic of the refractive index, $\kappa_0 = 2\pi/M_0$, $\kappa_a = 3.3/m_0$, $m_0$ and $M_0$ are the inner and outer scales of atmospheric turbulence.

To obtain statistical characteristics of intensity fluctuations, the Monte Carlo technique (statistical test method) is used. The calculations for different types of the beams were carried out for the same sample of 2400 random realizations of the sets of phase screens. After calculation of the complex amplitude of the field, the realization of the random field of intensity $I^{(j)}(\mathbf{r},z) = u^{(j)}(\mathbf{r},z)u^{(j)*}(\mathbf{r},z)$ and the squared intensity $[I^{(j)}(\mathbf{r},z)]^2$ at the end of the turbulent layer was calculated. The relative variance of intensity fluctuations (scintillation index) was calculated as

$$\sigma_I^2(\mathbf{r},z) = \frac{\langle I^2(\mathbf{r},z)\rangle}{\langle I(\mathbf{r},z)\rangle^2} - 1. \qquad (3)$$

The angular brackets in Eq. (3) denote the averaging over realizations, which was calculated in the standard way from the corresponding arrays of readings $\{I^{(j)}(\mathbf{r},z), [I^{(j)}(\mathbf{r},z)]^2, j=1,2,...,2400\}$. It was assumed that the turbulence at the path is statistically homogeneous $(C_n^2 = \text{const})$. The turbulence intensity was specified with the parameter $\beta_0^2 = 1.23 C_n^2 k^{7/6} z^{11/6}$.

The resultant dependences of the scintillation indices on turbulent conditions of propagation and the longitudinal coordinate of the observation point satisfying the condition $z/z_d = 1$ ($z_d = ka^2/2$ is the Rayleigh length) are shown in Figs. 1 and 2. In Fig. 1, curves 1 and 2 are the results of calculation of the scintillation indices at the beam axes. Curve 1 describes scintillations at the axis of the optical vortex (Laguerre-Gaussian mode of the $LG_0^1$ beam). Curve 2 was obtained for the *DH* beam. Curve 3 is borrowed from [22] (Fig. 6.26) and corresponds to the experimental results for the scintillation index of the narrow collimated Gaussian beam. It follows from Fig. 1 that the behavior of the scintillation indices of the optical vortex and the Gaussian beam in the zone of weak turbulence differs principally.

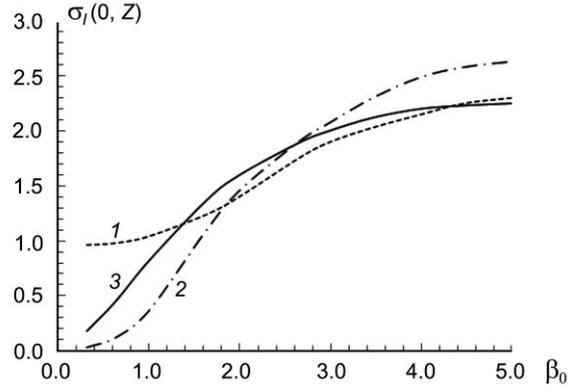

Fig. 1. Scintillations at the axis of the optical vortex: Laguerre-Gaussian beam $LG_0^1$ (curve 1) and *DH* beam (curve 2) and experimental results for the scintillation index of the narrow collimated Gaussian beam [22] (curve 3). Outer scale $M_0 = 20a$, $m_0 = 0.08a$.

The scintillation index of the optical vortex at the vortex axis increases sharply from zero to the value approximately equal to unity as the atmospheric turbulence "turns on," and then it increases smoothly at the further intensification of turbulence. Scintillations of the Gaussian beam increase smoothly from zero as the turbulence intensity increases. Scintillations of the *DH* beam also smoothly increase from zero, as those for the Gaussian beam.

Figure 2 shows the calculated scintillation indices in the transverse plane. Here, curves 1 and 2 correspond to the $LG_0^1$ and *DH* beams, while curve 3 is for the Gaussian beam. The calculations has been performed for the conditions of weak turbulence ($\beta_0^2 = 0.1$) and $z/z_d = 1$. It follows from the figure that scintillations of the optical vortex decrease down to those in the Gaussian and doughnut hole beams at a scale approximately equal to the effective source radius *a*. Then, the scintillation index demonstrates the approximately identical qualitative behavior in the range $\sigma_I^2(\mathbf{r}, z) = 1$ (curve 4) at the beam periphery and clear saturation to the level equal to unity at $r/a \gg 1$. The saturation of scintillations to the unit level seems to be more natural than the infinite increase of scintillations following from the estimates based on the Rytov theory [9]. It should be noted that the earlier numerical simulation of the Gaussian beam propagation [23] has also demonstrated the saturation of $\sigma_I^2(\mathbf{r}, z)$ at the beam periphery.

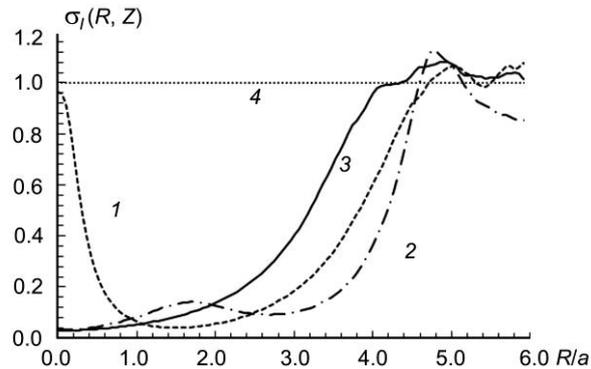

Fig. 2. Scintillation indices of optical vortex: Laguerre-Gaussian $LG_0^1$ beam (curve 1), *DH* beam (curve 2), and Gaussian beam (curve 3) as functions of the distance to the beam center; scintillation saturation level (4). $\beta_0^2 = 0.1$, $z/z_d = 1$. Outer scale $M_0 = 20a$, $m_0 = 0.08a$.

For analytical estimation of the scintillation index (3), we use the Kolmogorov spectrum [9]

$$\Phi_n(\boldsymbol{\kappa}_\perp,0) = 0{,}033 C_n^2 \kappa_\perp^{-11/3}.$$

The mean intensity at the axis of the optical vortex $\langle I(0,z)\rangle$ can be estimated by using the rigorous solution for $\Gamma_2(\mathbf{r}_1,\mathbf{r}_2,z) = \langle u(\mathbf{r}_1,z)u^*(\mathbf{r}_2,z)\rangle$ [24]. We use condition (1) the parameters corresponding to the optical vortex ($l=1$, $p=1$) and, considering turbulence as weak ($\beta_0^2 \ll 1$), obtain for the main term of the asymptotic series

$$\langle I(0,z)\rangle = 11.8 \sqrt[3]{4}\, \Gamma\left(\frac{11}{6}\right)\frac{\Phi}{c}\frac{1}{a^2}\frac{z_d}{z}\left(\frac{z z_d}{z^2+z_d^2}\right)^{11/6}. \tag{4}$$

In the calculation of $\langle I^2(0,z)\rangle$, we take into account that

$$\langle I^2(\mathbf{r},z)\rangle = \Gamma_4(\mathbf{r}_1,\mathbf{r}_2,\mathbf{r}_3,\mathbf{r}_4,z) = \langle u(\mathbf{r}_1,z)u^*(\mathbf{r}_2,z)u(\mathbf{r}_3,z)u^*(\mathbf{r}_4,z)\rangle_{\mathbf{r}_1=\mathbf{r}_2=\mathbf{r}_3=\mathbf{r}_4=\mathbf{r}},$$

and use the asymptotically rigorous method of equation solution for the first-order coherence function of the field $\Gamma_4(\mathbf{r}_1,\mathbf{r}_2,\mathbf{r}_3,\mathbf{r}_4,z)$ in the limiting case of weak turbulence [25]. It is to be recalled that this method assumes representation of $\Gamma_4(\mathbf{r}_{\underline{4}},z)$, $\mathbf{r}_{\underline{4}}=\mathbf{r}_{\underline{4}}(\mathbf{r}_1,\mathbf{r}_2,\mathbf{r}_3,\mathbf{r}_4)$ through the Green's function $G_4(\mathbf{r}_{\underline{4}},\mathbf{t}_{\underline{4}};z,0)$

$$\Gamma_4(\mathbf{r}_{\underline{4}},z) = \int \Gamma_4(\mathbf{t}_{\underline{4}},0) G_4(\mathbf{r}_{\underline{4}},\mathbf{t}_{\underline{4}};z,0)\, d\mathbf{t}_{\underline{4}} \tag{5}$$

and transition from the parabolic differential equation for $\langle G_4(\mathbf{r}_{\underline{4}},\mathbf{t}_{\underline{4}};z,0)\rangle$ to the integral equation, which can be written as a Neumann series

$$\langle G_4(\mathbf{r}_{\underline{4}},\mathbf{t}_{\underline{4}};z,0)\rangle = \sum_{j=0}^{\infty}(-1)^j \langle G_4(\mathbf{r}_{\underline{4}},\mathbf{t}_{\underline{4}};z,0)\rangle_j. \tag{6}$$

The corresponding iteration series for $\Gamma_4(\mathbf{r}_{\underline{4}},z)$ has the form

$$\Gamma_4(\mathbf{r}_{\underline{4}},z) = \sum_{j=0}^{\infty}(-1)^j \Gamma_{4j}(\mathbf{r}_{\underline{4}},z).$$

In the zone of weak turbulence, the equation corresponding to the homogeneous medium [25] can be taken as $\langle G_4(\mathbf{r}_{\underline{4}},\mathbf{t}_{\underline{4}};z,0)\rangle_0$. Then

$$\Gamma_{40}(\mathbf{r}_{\underline{4}},z) = u_0(\mathbf{r}_1,z)u_0^*(\mathbf{r}_2,z)u_0(\mathbf{r}_3,z)u_0^*(\mathbf{r}_4,z), \tag{7}$$

where $u_0(\mathbf{r}_1,z)$ is the complex amplitude of the field in the homogeneous medium at a distance $z$ from the source. It is obvious that $\Gamma_{40}(0,z)=0$ at the axis of the optical vortex. The first iteration allows us to find that $\Gamma_{41}(0,z)=0$ as well, so that

$$\Gamma_4(0,z) = \sum_{j=2}^{\infty}(-1)^j \Gamma_{4j}(0,z). \tag{8}$$

Estimating $\Gamma_{42}(0,z)$, we obtain

$$\Gamma_{42}(0,z) \cong 2\langle I(0,z)\rangle^2,$$

where $\langle I(0,z)\rangle$ is described by Eq. (4).

With allowance for the terms of asymptotic series (8) following the term $\Gamma_{42}(0,z)$, we obtain

$$\langle I^2(0,z)\rangle \cong 2\langle I(0,z)\rangle^2 + O(\beta_0^6), \quad \beta_0^2 \ll 1. \tag{9}$$

Then, according to Eqs. (3) and (9), we have the following estimate for the scintillation index

$$\sigma_I^2(0,z) = 1 + O(\beta_0^2), \quad \beta_0^2 \ll 1. \tag{10}$$

This estimate indicates the stepwise character of the increase of scintillations in the optical vortex at intensification of atmospheric turbulence and corresponds to the results of $\sigma_I^2(0,z)$ calculation in the numerical experiment (curves 1 in Figs. 1 and 2.).

It should be noted that the characteristic $\sigma_I(0,z)$ can be considered as a speckle contrast [26], and fulfillment of the condition $\sigma_I(0,z) \approx 1$ is one of the signs of the fully developed speckle field, which is formed in the cross section of the $LG_0^1$ beam as a regular structure and occupies a random position in the transverse plane due to random inhomogeneities of the medium.

It is obvious that the approximate equality $\sigma_I(0,z) \approx 1$ is a consequence of the deep spatial modulation of the regular distribution $I_0(\mathbf{r},z)$ provided by the presence of the zero intensity value. Actually, using the Michelson contrast [27] equal to the ratio of the difference between the maximal and minimal values of intensity in the beam cross section to the sum of the maximal and minimal values, for estimation of the depth of spatial modulation of the intensity field, we obtain eth contrast equal to 1 (100%) owing to the fact that $I_0(0,z) = 0$. The 100% contrast will keep its value not only for the beam in the homogeneous medium, but also in every particular realization $I(\mathbf{r},z)$ of the $LG_0^1$ beam propagating in the weakly turbulent atmosphere. Under these conditions, the zero intensity being, according to [28], a stable structure keeps in the beam intensity distribution. It should be recalled that when the condition $\beta_0^2 \gg 1$ is satisfied, we observe the "mode of saturated scintillations," which is characterized by eth appearance of natural speckle structures (wave front dislocations) [29,30]. It should be noted that the regular intensity distribution of the $DH$ beam also has the deep spatial modulation in the plane $z = 0$. However, owing to diffraction, the Michelson contrast for this beam decreases with the increase of the evolutionary variable $z$ and becomes equal to the corresponding value for the Gaussian beam [19].

Thus, we have performed the numerical and analytical estimation of the scintillation index for the $LG_0^1$, $DH$, and Gaussian beams in the randomly inhomogeneous atmosphere. It has been found that the scintillation index in the $LG_0^1$ beam increase stepwise from zero to unity at the turbulence development at the path in contrast to scintillations of the two other beams. This occurs owing to the initial deep spatial modulation of the transverse intensity distribution, which becomes random due to random inhomogeneities of the medium. The dynamics of scintillations in the $DH$ beam obeys the same qualitative regularities as scintillations in the Gaussian beam do [9]. The strong inhomogeneity of scintillations in the cross section of the $LG_0^1$ beam has been observed. It has been found that at the beam periphery the scintillation index of all the studied beams saturates to the unit level.